\documentclass[a4paper]{PoS}

\def\dd{{\mathrm{d}}}

\title{On the light-front wave functions of quarkonia}

\ShortTitle{On the light-front wave functions of quarkonia}

\author{\speaker{Pieter Maris}, Shaoyang Jia, Meijian Li, Yang Li, Shuo Tang and James P. Vary\\
  Dept. of Physics and Astronomy, Iowa State University, Ames, IA 50011, USA\\
  \email{pmaris@iastate.edu},
  \email{sjia@iastate.edu},
  \email{meijianl@iastate.edu},
  \email{leeyoung@iastate.edu},
  \email{tang@iastate.edu},
  \email{jvary@iastate.edu}}

\abstract{
  The light-front wave functions of hadrons allow us to calculate
  a wide range of physical observables; however, the wave functions
  themselves cannot be measured.  We discuss recent results for
  quarkonia obtained in basis light-front quantization using an
  effective Hamiltonian with a confining model in both the transverse
  and longitudinal directions and with explicit one-gluon exchange.
  In particular, we focus on the numerical convergence of the basis
  expansion, as well as the asymptotic behavior of the light-front
  wave functions.  We also illustrate that, for mesons with unequal
  quark masses, the maxima of the light-front wave functions depend
  in a non-trivial way on the valence quark-mass difference.}

\FullConference{Light Cone 2019 - QCD on the light cone: from hadrons to heavy ions - LC2019\\
		16-20 September 2019\\
		Ecole Polytechnique, Palaiseau, France}

\begin{document}

\section{Confining model for quarkonia}

We use a model for the confining potential between a quark and an
anti-quark in a color-singlet in both the longitudinal and the
transverse direction, as described in
Refs.~\cite{Li:2015zda,Li:2017mlw}.  The short-range, high-momentum
physics is dominated by one-gluon exchange, which we add to the
confining potential.  Thus the effective light-front Hamiltonian for a
quark and an anti-quark in a flavor-singlet configuration becomes
\begin{eqnarray}
  H_\mathrm{eff} &=&
  \frac{\vec k^2_\perp + m_q^2}{x} + \frac{\vec k^2_\perp+m_{\bar q}^2}{1-x}
  + \kappa^4 \vec \zeta_\perp^2 - \frac{\kappa^4}{(m_q+m_{\bar q})^2} \partial_x\left( x(1-x) \partial_x \right)
  \nonumber \\
  & &  - C_F \frac{4\pi\alpha_s(Q^2)}{Q^2}\bar u_{s'}(k')\gamma_\mu u_s(k) \bar v_{\bar s}(\bar k) \gamma^\mu v_{\bar s'}(\bar k') \,,
  \label{Eq:Heff}
\end{eqnarray}
where $m_q$ and $m_{\bar q}$ are the masses of the quark and anti-quark,
$\vec \zeta_\perp \equiv \sqrt{x(1-x)} \vec r_\perp$ is Brodsky
and de T\'eramond's holographic variable~\cite{Brodsky:2014yha},
$\kappa$ is the strength of the confinement, and the longitudinal confinement
is described by
$\partial_x f(x, \vec\zeta_\perp) = \partial f(x, \vec \zeta_\perp)/\partial x|_{\vec\zeta}$.
The second line corresponds to the one-gluon exchange, with
$C_F = (N_c^2-1)/(2N_c)=4/3$ the color factor for the color singlet
state, and $Q^2 = -q^2 > 0$ is the 4-momentum squared carried by the
exchanged gluon~\cite{Li:2017mlw}.

The mass spectrum and corresponding light-front wave functions (LFWF)
are obtained by diagonalizing the effective light-front Hamiltonian
operator (\ref{Eq:Heff})
\begin{eqnarray}
  H_\mathrm{eff} |\psi_h(P, j, m_j)\rangle &=& M^2_h |\psi_h(P, j, m_j)\rangle \,,
  \label{Eq:LFHam}
\end{eqnarray}
where $P=(P^-, P^+, \vec P_\perp)$ is the 4-momentum of the meson, and
$j$ and $m_j$ are the particle's total angular momentum and magnetic
projection, respectively.  In the leading $q\bar{q}$ Fock space we have
\begin{eqnarray}
  |\psi_h(P, m_j)\rangle & \approx & 
  \sum_{s, \bar s}\int_0^1\frac{\dd x}{2x(1-x)}
  \int \frac{\dd^2 k_\perp}{(2\pi)^3}
  \; \psi^{(m_j)}_{s\bar s/h}(\vec k_\perp, x)
  \nonumber \\
  && \times \frac{1}{\sqrt{N_c}}\sum_{i=1}^{N_c}
  b^\dagger_{s{}i}\left(xP^+, \vec k_\perp+x\vec P_\perp\right)
  d^\dagger_{\bar s{}i}\left((1-x)P^+,-\vec k_\perp+(1-x)\vec P_\perp\right) |0\rangle \,,
\end{eqnarray}
with $b^\dagger$ and $d^\dagger$ the quark and anti-quark creation
operators and $\psi^{(m_j)}_{s\bar s/h}(\vec k_\perp, x)$ the valence
space LFWF with $s$ and $\bar s$ the spin of the quark and antiquark,
properly normalized to
\begin{eqnarray}
  \sum_{s, \bar s} \int_0^1\frac{\dd x}{2x(1-x)}
  \int \frac{\dd^2  k_\perp}{(2\pi)^3}  \;
  \psi^{(m_j')*}_{s \bar s/h'}(\vec k_\perp, x) \,
  \psi^{(m_j)}_{s \bar s/h}(\vec k_\perp, x)  &=& \delta_{hh'}\delta_{m_j,m_j'} \,.
\end{eqnarray}

Without the one-gluon exchange, this model can be solved analytically,
and the LFWFs can be expressed as a product of a 2-dimensional
harmonic oscillator (HO) function $\phi_{nm}$ with strength parameter
$\kappa$ and a Jacobi polynomial times power-law factors $\chi_l$ in
the longitudinal direction~\cite{Li:2015zda}.  These analytic
solutions form a convenient and complete basis for expanding the LFWFs
\begin{eqnarray}
  \psi_{ss'/h}(\vec k_\perp, x) = \sum_{n, m, l}
  c_{ss'/h}^{nml} \; \phi_{nm}\left(\vec k_\perp/\sqrt{x(1-x)}\right) \, \chi_l(x) \,.
\end{eqnarray}
By expressing our effective Hamiltonian Eq.~(\ref{Eq:Heff}) in this
basis, Eq.~(\ref{Eq:LFHam}) becomes a matrix equation for the
coefficients $c_{ss'/h}^{nml}$, which we diagonalize numerically.
In the limit of a complete (but infinite-dimensional) basis, this
gives us the exact LFWFs for this Hamiltonian in the leading Fock
space.  This model has been used to compute a number of observables as
such as radiative decays and form factors~\cite{Li:2018uif,Li:2020wrn},
well as in diffractive vector meson production~\cite{Chen:2018vdw},
showing reasonable agreement with the available experimental data.

\section{Numerical convergence}

The precision of our numerical calculations depends both on the number
of basis states that we keep in our expansion, and on the precision of
the evaluation of the Hamiltonian matrix elements in this basis.  The
number of basis states is controlled by the truncation parameters
$N_{\max}$ and $L_{\max}$ for the transverse and longitudinal
directions, respectively; for simplicity, we keep $N_{\max}=L_{\max}$.
The matrix elements are evaluated numerically, using $n_x$ and $n_k$
integration points for the $x$ and $k_\perp$ integrations,
respectively; at a minimum we keep $n_x \ge L_{\max}$ and $n_k \ge
N_{\max}$.

\subsection{Meson mass and electroweak decay constants}

\begin{figure}[b]
  \includegraphics[width=.95\textwidth]{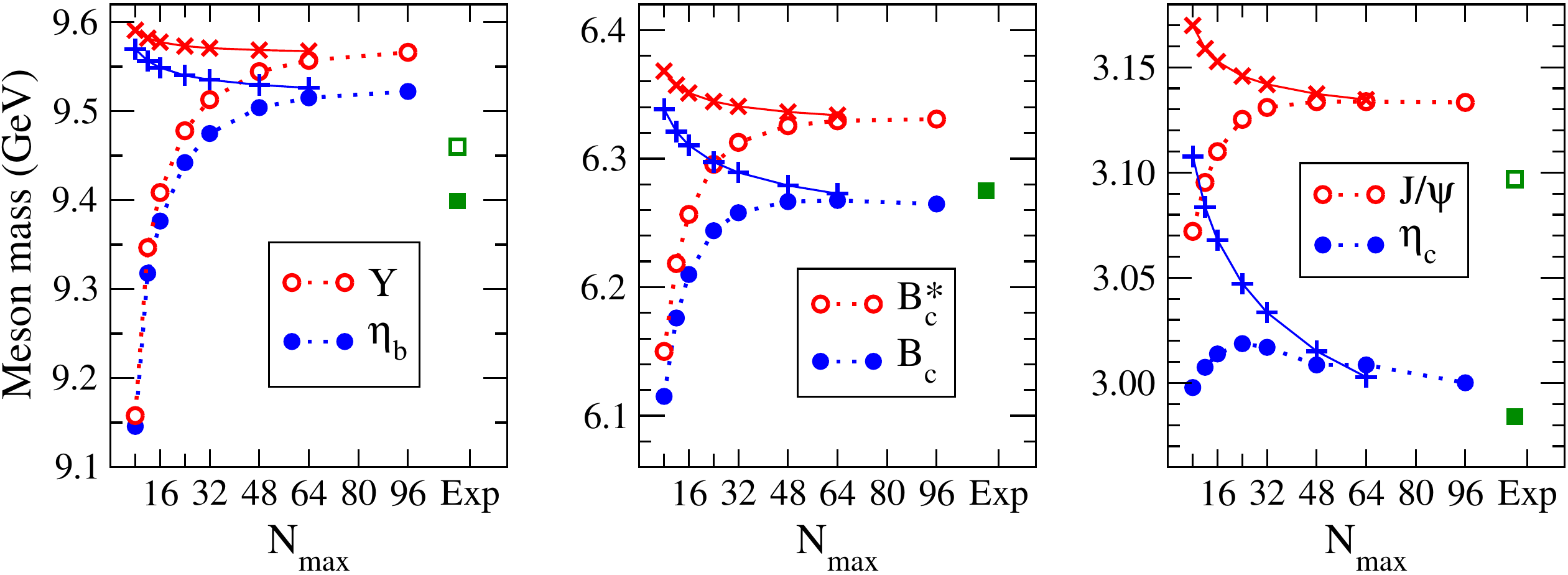}
  \includegraphics[width=.95\textwidth]{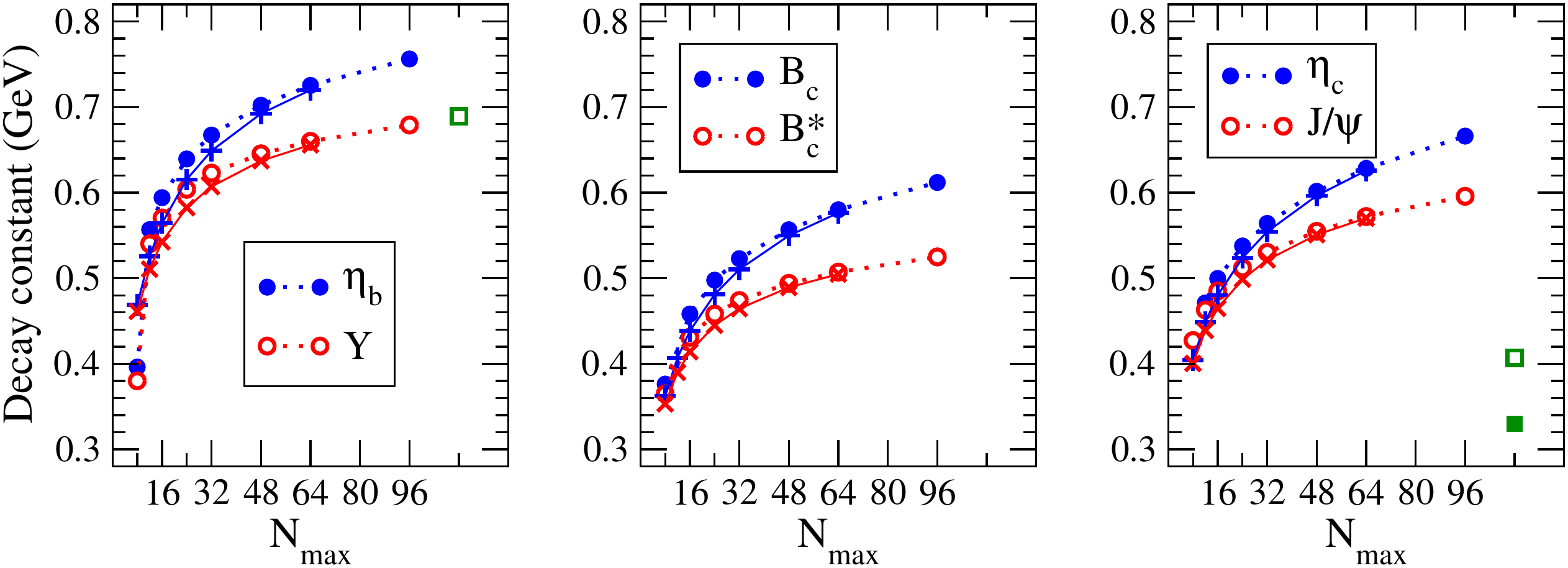}
  \caption{
    The meson mass (top) and decay constant (bottom)
    as function of the basis truncation $N_{\max}$ with
    $n_x = n_k = 2 N_{\max}$ (open and closed circles)
    and with $n_x = 192$ and $n_k = 96$ (plusses and crosses),
    and experimental values from Ref.~\cite{Tanabashi:2018oca}
    (open and closed green squares). 
    \label{Fig:Convergence} }
\end{figure}

The dependence of the masses and decay constants on the basis
truncation parameter $N_{{\max}}$ is shown
Fig.~\ref{Fig:Convergence} for the lowest pseudoscalar and vector
$Q\bar{q}$ states.  The meson masses appear to converge reasonably
well, both with $N_{{\max}}$ and with the number of integration points
$n_x$ and $n_k$.  They do not necessarily converge to the physical
values -- the parameters we use here were fitted in
Ref.~\cite{Li:2017mlw} at $N_{\max} = L_{\max} = 32$ and
$n_x = n_k = 64$ to the lowest 14 and 8 states of bottomonium and
charmonium respectively.  Note that as we increase $N_{\max}$ while
keeping $n_x$ and $n_k$ fixed, the obtained masses decrease
monotonically, in agreement with the variational principle.

On the other hand, the obtained decay constants seem to be (almost)
independent of $n_x$ and $n_k$, but they depend strongly on the basis
truncation parameter $N_{\max}$.  This suggest that the decay
constants (in contrast to the masses) are sensitive to the
high-momentum behavior of the LFWF.  Indeed, it is known from e.g. the
covariant Dyson--Schwinger approach that the integral for the
pseudoscalar decay constant over the Bethe--Salpeter amplitude $\chi_{\rm PS}(k, P)$ 
\begin{eqnarray}
  f_{\rm PS} &=& \frac{Z_2}{m^2_{\rm PS}}
  \int \frac{\dd^4k}{(2\pi)^4} {\rm Tr}[ \chi_{\rm PS}(k, P) \gamma_5  /\!\!\!\!\!P ] 
\end{eqnarray}
is potentially
divergent.  This divergence is absorbed by the wave function
renormalization constant $Z_2$, rendering a finite result for the
physical decay constants~\cite{Maris:1997hd}.  In
Ref.~\cite{Li:2017mlw} we have therefore truncated the corresponding
integral in the transverse direction of the light-front formalism
\begin{eqnarray}
  f_{\rm PS} & = & \sqrt{2 N_c}
  \int_0^1 \frac{\dd x}{\sqrt{x(1-x)}} \;
  \int^\mu \frac{\dd^2k_\perp}{(2\pi)^3}  \;
  \psi_{\uparrow\downarrow-\downarrow\uparrow}(\vec{k}_\perp, x)
\end{eqnarray}  
at an appropriate UV mass scale $\mu \approx 1.7 m_q$.  In order
to address the question of the convergence of the decay constant, we
now turn our attention to the asymptotic behavior of the LFWFs.

\subsection{Asymptotics of Light-Front wave functions}

Understanding the ultraviolet (UV) asymptotics of the LFWF as
$k_\perp \to \infty$ is crucial for the proper evaluation
(with consistent regularization and renormalization as necessary)
of observables such as decay constants and elastic and transition form
factors.  The infrared behavior of the LFWF is dominated by our model
for the confining interaction as well as the behavior of the running
coupling at small momenta $Q^2$.  On the other hand, we expect the UV
behavior to be dominated by perturbative QCD, and in particular the
large-$Q^2$ behavior of the one-gluon exchange.

\begin{figure}[b]
  \includegraphics[width=.45\textwidth]{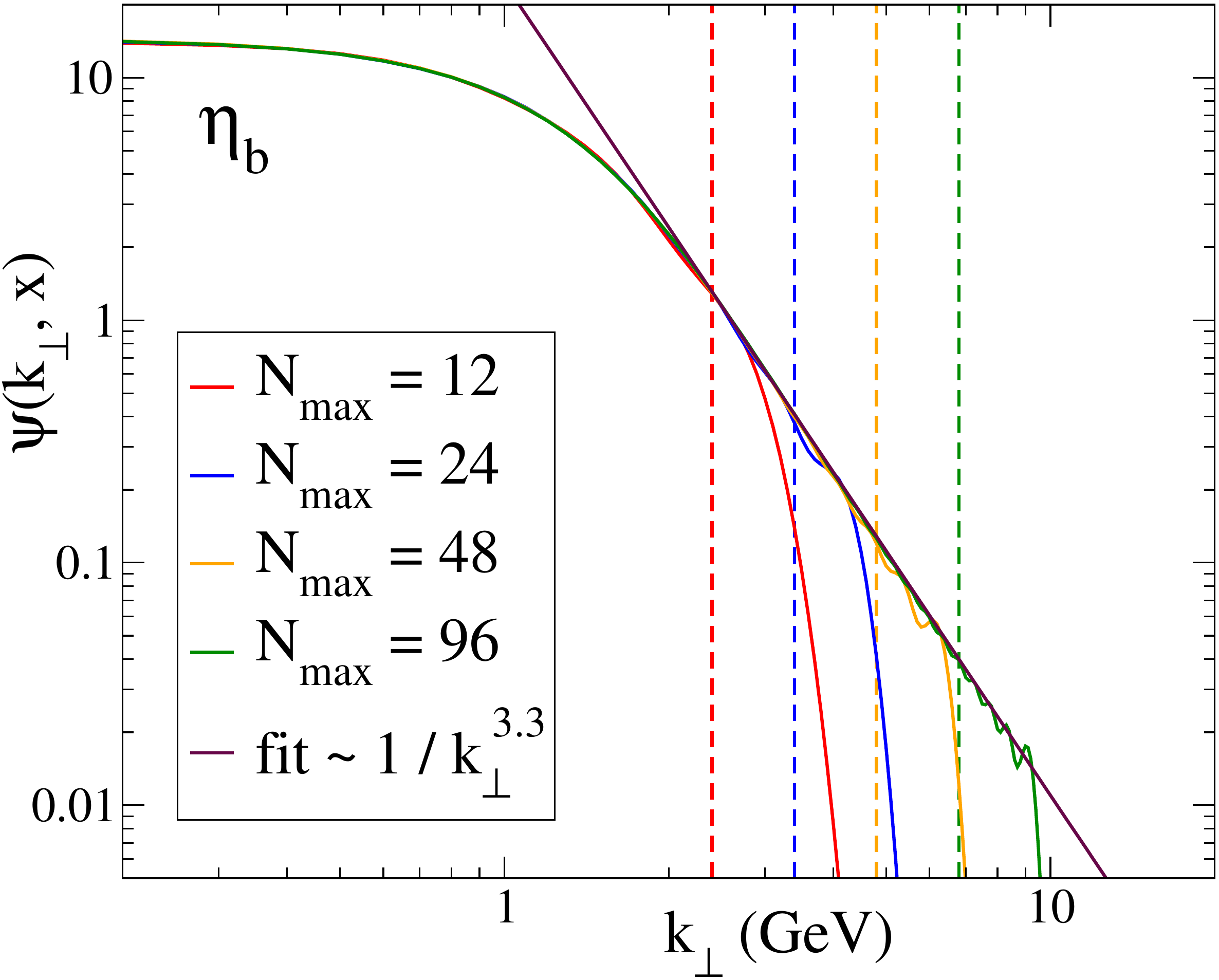}
  \qquad \includegraphics[width=.45\textwidth]{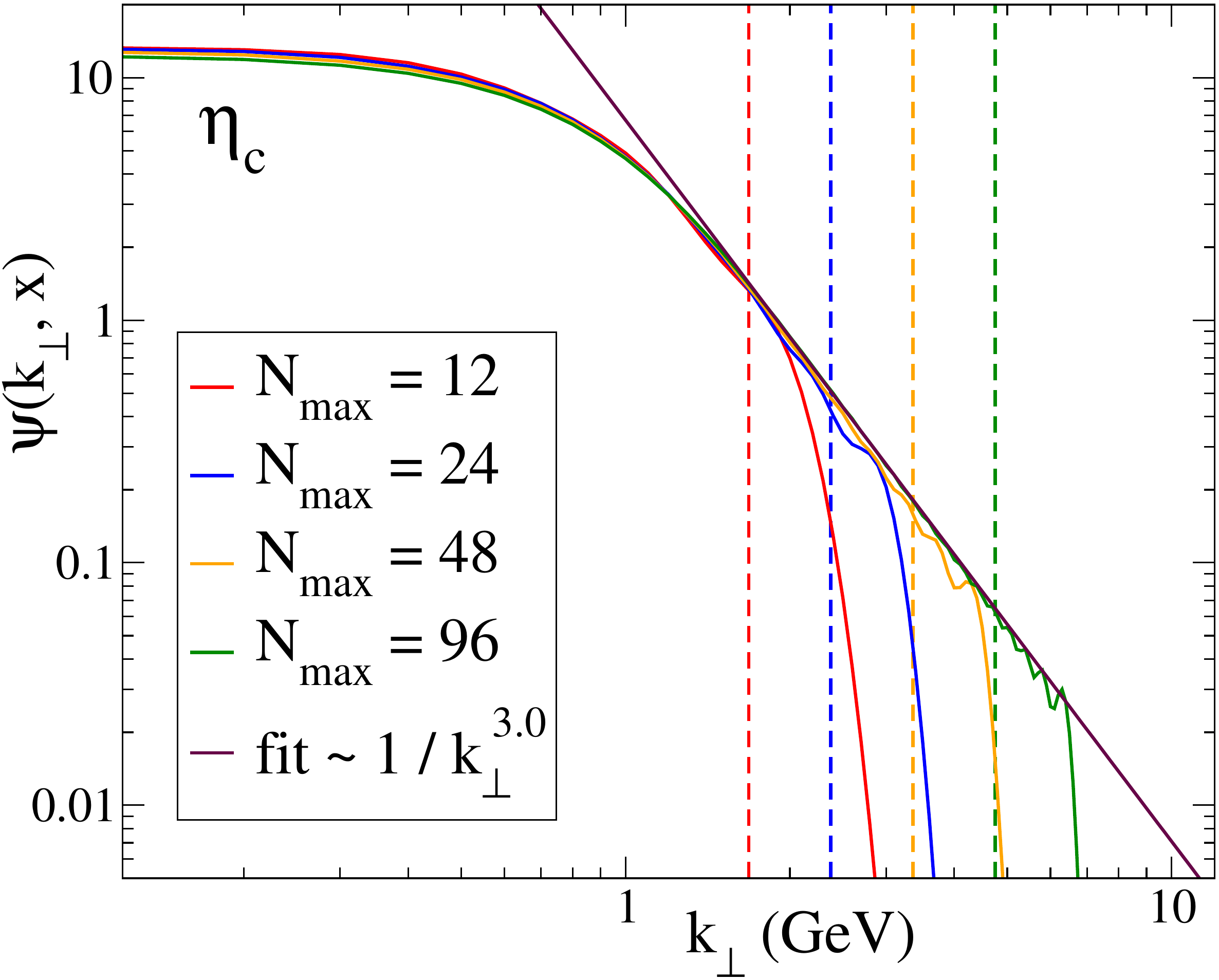}
  \caption{
    Dominant component $\psi_{\uparrow\downarrow-\downarrow\uparrow}(k_\perp, x)$
    of the LFWF for $\eta_b$ (left) and $\eta_c$ (right) as function
    of $k_\perp$ at $x=\frac{1}{2}$ on a log-log scale.  The vertical
    dashed lines indicate the UV truncation scale $\Lambda$ defined in
    the text.
    \label{Fig:LFWF_as} }
\end{figure}
Within our finite basis calculations, the asymptotics can
only be well-represented up to the effective UV truncation scale,
$\Lambda = \kappa \sqrt{x(1-x) \, N_{\max}}$, in our transverse basis
functions.  Indeed, Fig.~\ref{Fig:LFWF_as} shows that below this
scale $\Lambda$, the LFWFs of the $\eta_b$ and $\eta_c$ are
(almost) independent of the truncation parameter $N_{\max}$,
but for $k_\perp > \Lambda$ the LFWF starts to oscillate and falls off
like a gaussian, as one would expect.  For the vector mesons we obtain
the same asymptotic behavior.  Our calculations suggest that the
LFWFs fall off like $1/k_\perp^3$ or even faster (possibly with a
logarithmic correction), but an expansion in HO basis functions in
the transverse direction is far from the ideal computational method
for an accurate determination of the asymptotic behavior of the LFWFs.

\subsection{Asymptotics of the transverse Distribution Amplitude}

Analogous to the (longitudinal) Distribution Amplitudes, which are
obtained from the LFWF by integrating over the transverse momenta, one
can define Transverse Distribution Amplitudes (TDA) by integrating over
$x$
\begin{eqnarray}
  \phi_\perp(\vec{k}_\perp) & = & \sqrt{2 N_c}
  \int_0^1 \frac{\dd x}{\sqrt{x(1-x)}} \;
  \psi_{\uparrow\downarrow-\downarrow\uparrow}(\vec{k}_\perp, x) \,,
\end{eqnarray}  
normalized here such that the integral over $\vec{k}_\perp$ gives the
decay constant
\begin{eqnarray}
  f_{\rm PS} & = & \int \frac{\dd^2 \vec{k}_\perp}{(2\pi)^3}
  \; \phi_\perp(\vec{k}_\perp) \,.
\end{eqnarray}  
\begin{figure}[b]
  \includegraphics[width=.45\textwidth]{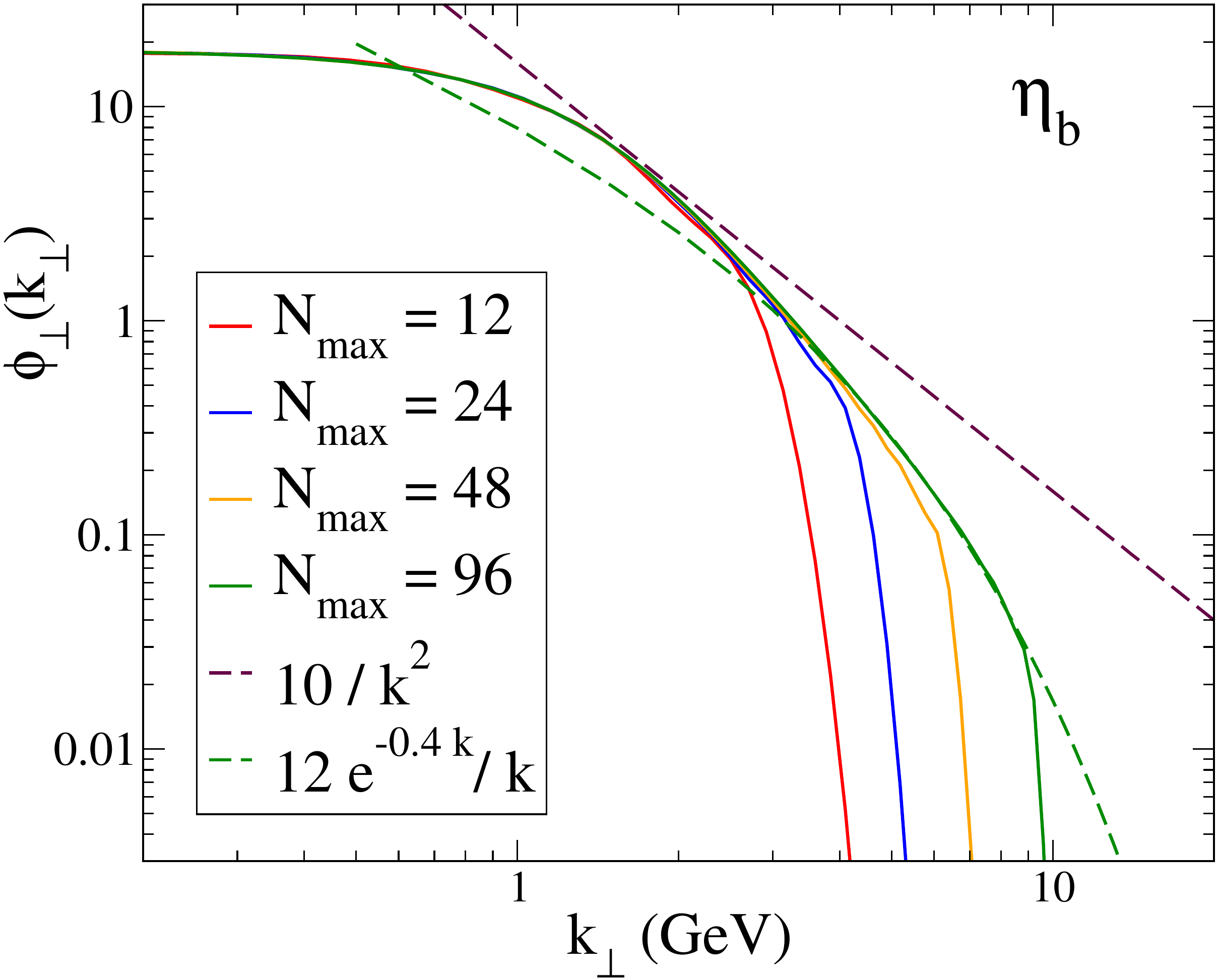}
  \qquad \includegraphics[width=.45\textwidth]{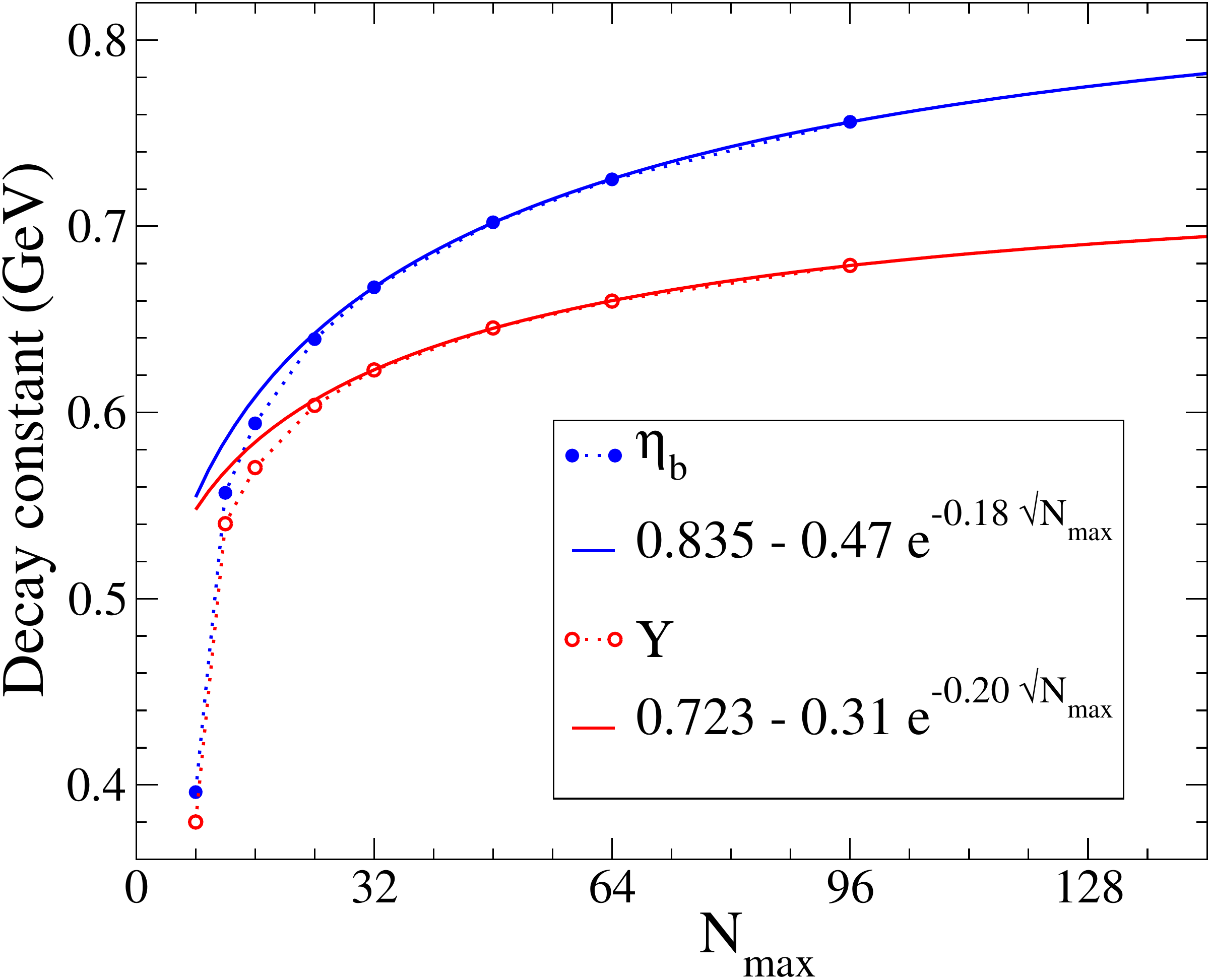}
  \caption{
    Transverse DA for $\eta_b$ (left) and extrapolation of the decay
    constants (right).
    \label{Fig:TDA_as} }
\end{figure}
The left panel of Fig.~\ref{Fig:TDA_as} shows $\phi_\perp(k_\perp)$ of
$\eta_b$ for several different basis truncation parameters $N_{\max}$.
This shows good convergence up to a scale proportional to
$\sqrt{N_{\max}}$.  This figure clearly shows that the TDA in this
model falls off faster than $1/k_\perp^2$.  Our best fit for the
asymptotic behavior is
\begin{eqnarray}
  k_\perp \; \phi_\perp(\vec{k}_\perp) & \sim & {\rm e}^{-c\, k_\perp} \,,
\end{eqnarray}
which suggests that the integral for the decay constant is finite,
even in the limit of a complete basis.  As indicated before, we may
need alternative computational tools to accurately determine the
asymptotic behavior of the TDA, and whether this behavior follows from
the one-gluon exchange, or from the one-gluon exchange in combination
with our specific model for the transverse and longitudinal
confinement.

Inspired by this behavior, we extrapolate the decay constants by a fit
of our finite basis results
\begin{eqnarray}
  f(N_{\max}) &=& f(N_{\max}=\infty) + a {\rm e}^{-c\, \sqrt{N_{\max}}} \,,
\end{eqnarray}
as is shown in the right panel of Fig.~\ref{Fig:TDA_as} for the
$\eta_b$ and $\Upsilon$.  This extrapolation also works quite well for
the decay constants of the $\eta_c$, $J/\psi$, and $B_c$.  Of course,
these extrapolated decay constants are significantly larger than those
reported in Refs.~\cite{Li:2017mlw,Tang:2018myz}, which were truncated
at a finite scale.  

\section{LFWF of unequal-mass heavy mesons}

\begin{figure}[b]
  \includegraphics[width=.45\textwidth]{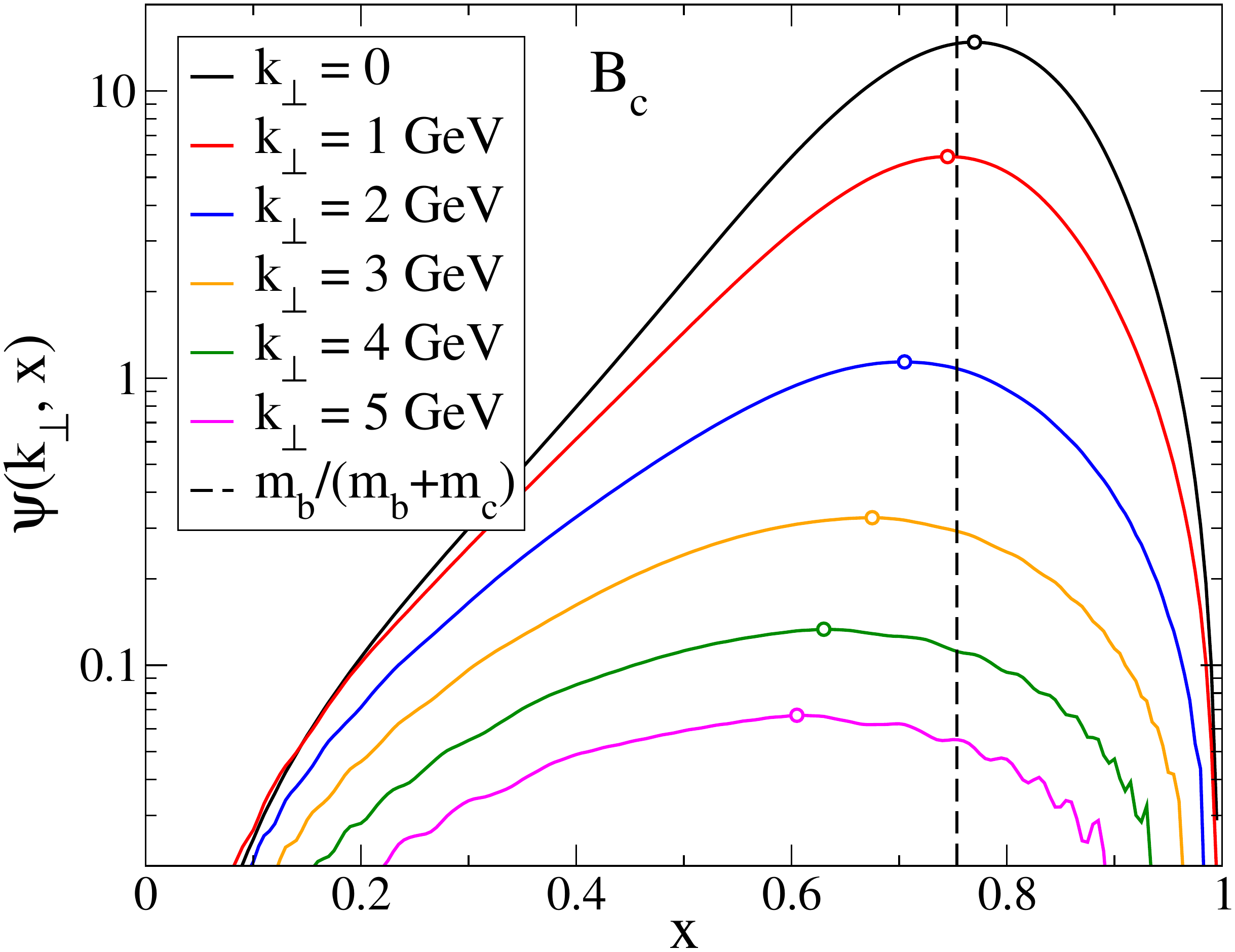}
  \qquad \includegraphics[width=.45\textwidth]{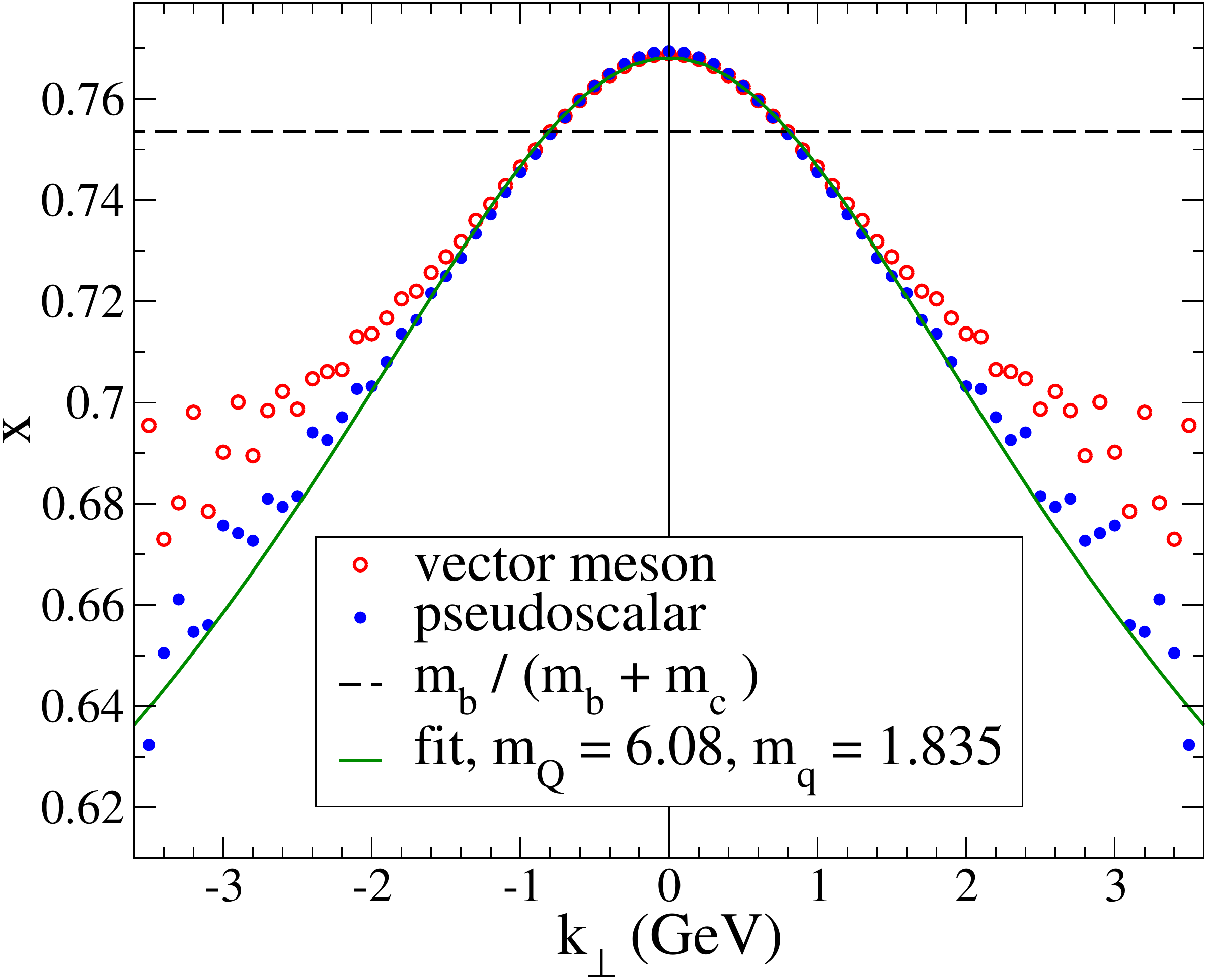}
  \caption{
    Dominant component $\psi_{\uparrow\downarrow-\downarrow\uparrow}(x, k_\perp)$
    of the LFWF for $B_c$ at fixed $k_\perp$ (left) and
    position of the maximum in $x$ in the LFWF at fixed
    $k_\perp$ as a function of $k_\perp$ (right),
    with the fit Eq.~(\ref{Eq:cond_peak_rho}).
    \label{Fig:Bc_LFWF} }
\end{figure}

For equal-mass constituents, the LFWF $\psi(\vec k_\perp, x)$ has its
maximum value at $x = \frac{1}{2}$ for all $k_\perp$, but for
systems with valence quarks of unequal masses such as $B_c$, this is
not the case.  Nonperturbatively, one expects this maximum to occur at
$\frac{m_Q}{m_Q+m_q}$; however, the left panel of
Fig.~\ref{Fig:Bc_LFWF} clearly shows that this maximum depends on
$k_\perp$.  At $k_\perp = 0$, the peak-position (indicated by an open
circle) is at $x > \frac{m_b}{m_b+m_c}$, but for increasing $k_\perp$,
this maximum occurs at lower values of $x$.  In the limit
$k_\perp\rightarrow \infty$ the effect of the unequal masses becomes
negligible, and the maximum position approaches $x=\frac{1}{2}$.

With the LFWF interpreted as the probability amplitude, its peak
coincides with the maximum of the probability distribution for finding
the system with in a given momentum configuration.  This
momentum-space probability density can be alternatively calculated
using the light-front parton gas model~\cite{Jia:2018hxd}.
Specifically, the joint probability distribution in $k_\perp$ and $x$
is given by
\begin{eqnarray}
  \rho(k_\perp,x) &\sim& \delta\left(\frac{k_{\perp}^2+m_Q^2}{x}+\frac{k_{\perp}^2+m_q^2}{1-x}-u\right)\,,
  \label{eq:rho}
\end{eqnarray}
where the quantity $u$ is the available thermal energy for the
relative motion of the valence quarks, after averaging interactions
other than the light-front kinetic energy.  The parameters $m_Q$ and
$m_q$ need not to be identical to those in the light-front Hamiltonian
because of this averaging -- one can think of these parameters as
`effective masses'.
	
Notice that with fixed $x$, the peak of the distribution always
locates at $k_\perp=0$.   The peak of the LFWF with fixed
$k_\perp$ is given by the $x$ which makes the partial derivative of
Eq.~(\ref{eq:rho}) with respect to $x$ vanish.  The derivative of the
$\delta$-function can be understood by approximating it with a narrow
gaussian function.  Explicitly, we obtain for the peak-location in $x$ at fixed $k_\perp$
\begin{eqnarray}
  x &=&\left( 1+\sqrt{\frac{k_\perp^2+m_q^2}{k_\perp^2+m_Q^2}}\right)^{-1} \,,
  \label{Eq:cond_peak_rho}
\end{eqnarray}
which indeed describes the peak-position in our model, see the right
panel of Fig.~\ref{Fig:Bc_LFWF}, with the same effective masses $m_Q$
and $m_q$ for the lowest pseusoscalar and vector states.  As a
consequence, different light-front observables such as the
Distribution Amplitude (DA) and the Parton Distribution Function (PDF)
have their maximum at different values of $x$.  This effect is more
pronounced as the mass difference between the quarks becomes
larger~\cite{Tang:2019gvn}.

\section*{Acknowledgements}
This work was supported by the US Department of Energy under Grants
No. DE-FG02-87ER40371 and No. DE-SC0018223 (SciDAC-4/NUCLEI).
Compuutational resources were provided by the National Energy Research
Scientific Computing Center (NERSC), which is a US Department of
Energy Office of Science user facility, supported under Contract
No. DE-AC02-05CH11231.

\end{document}